\documentclass[useAMS,usenatbib,usegraphicx]{mn2e}
\usepackage{epsfig}
\usepackage{amsmath} 
\usepackage{rotating}           
\usepackage{color}     
\usepackage{graphicx}
\usepackage{times}
\usepackage{upgreek} 
\usepackage{amssymb} 

\def\kms{km ${\rm s}^{-1}$}

\def\scm  {$\hbox{{\rm cm}}^{-2}$}    
\def \AL {$\alpha $}     
\def \HI {H{\sc \,i}}
\def \WpHz {W Hz$^{-1}$}

\def\lapp{\ifmmode\stackrel{<}{_{\sim}}\else$\stackrel{<}{_{\sim}}$\fi}
\def\gapp{\ifmmode\stackrel{>}{_{\sim}}\else$\stackrel{>}{_{\sim}}$\fi}
\title[Ionisation in redshifted radio sources]{Ionisation of the atomic gas in redshifted radio sources}
\author[S. J. Curran,  et al.]{S. J. Curran$^{1}$\thanks{Stephen.Curran@vuw.ac.nz},  R. W. Hunstead$^{2}$,  H. M. Johnston$^{2}$, M. T. Whiting$^{3}$, E. M. Sadler$^{2,3}$,  
\newauthor  J. R. Allison$^{4}$ and  R. Athreya$^{5}$\\
$^{1}$School of Chemical and Physical Sciences, Victoria University of Wellington, PO Box 600, Wellington 6140, New Zealand\\
$^{2}$Sydney Institute for Astronomy, School of Physics, The University of Sydney, NSW 2006, Australia\\
$^{3}$CSIRO Astronomy and Space Science, PO Box 76, Epping NSW 1710, Australia\\
$^{4}$Sub-department of Astrophysics; Department of Physics, University of Oxford, Denys Wilkinson Building, Keble Road, Oxford, OX1 3RH, U.K.\\
$^{5}$Indian Institute of Science Education and Research, 900, NCL Innovation Park, Dr Homi Bhabha Road Pune, Maharashtra 411008, India}

\begin{document}

\date{Accepted ---. Received ---; in original form ---}

\pagerange{\pageref{firstpage}--\pageref{lastpage}} \pubyear{2019}

\maketitle

\label{firstpage}

\begin{abstract}
  We report the results of a survey for \HI\ 21-cm absorption at $z\lapp0.4$ in a new sample of radio sources with the
  Giant Metrewave Radio Telescope. Of the 11 sources for which there are good data, we obtain zero detections, where four
  are expected upon accounting for the ionising photon rates and sensitivity. Adding these to the previously
  published values, we confirm that the non-detection of 21-cm absorption in active sources at high redshift is due to
  photo-ionisation of the gas rather than excitation by 21-cm photons (significant at $6.09\sigma$ and $2.90\sigma$,
  respectively).  We also confirm a strong correlation between the absorption strength and the reddening of the source,
  suggesting that dust plays a significant role in shielding the gas from the ambient ultra-violet field.  An
  anti-correlation between the 21-cm detection rate and the radio turnover frequency is also found, which runs contrary
  to what is expected on the basis that the higher the turnover frequency, the more compact the
  source.  It is, however, consistent with the hypothesis that the turnover frequency is related to the electron density,
  supported by a correlation between the turnover frequency and ionising photon rate.
\end{abstract} 


\section{Introduction}
\label{intro}

Cold neutral gas, the reservoir for star formation throughout the Universe, is traced through absorption
of the background  21-cm continuum by  neutral hydrogen (\HI). Beyond the Milky Way, this is either
detected in quiescent galaxies, which intervene the sight line to more distant radio sources, or within
the host of the continuum source itself. In the former {\em intervening} systems,
absorption via the Lyman-\AL\ transition of \HI\ is often also detected (and usually a prerequisite for 
21-cm searches), while this is not the case for the latter {\em associated} systems. 
This  is due to the  ``proximity effect'', where the absorption by neutral atomic gas (or its tracers, e.g.
Mg{\sc ii}) becomes sparser as the absorption redshift approaches that of the background source, due to the 
higher ionising ($\lambda_{\rm rest}\lapp912$ \AA) fluxes (e.g. \citealt{wcs81,bdo88,wkw+08,jcs18}).

A similar effect has been seen for the 21-cm transition, where associated \HI\ 21-cm absorption has never been detected
in radio sources above a ``critical UV luminosity'' of $L_{\rm UV}\sim10^{23}$ \WpHz\ \citep{cww+08}, a result which has
been confirmed several times since
(\citealt{cwm+10,cwsb12,cwt+12,caw+16,chj+17,cwa+17,ace+12,gmmo14,akk16,ak17,gdb+15}).\footnote{\citet{gdb+15}, which is
  still in submission, report 0 new detections of \HI\ 21-cm absorption out of 89 new searches over $0.02 < z < 3.8$
  (see \citealt{gd11}).} This luminosity corresponds to an ionising photon rate of $Q_\text{\HI}\equiv
\int^{\infty}_{\nu^{\prime}}({L_{\nu}}/{h\nu})\,d{\nu}\approx3\times10^{56}$~s$^{-1}$ (where
$\nu^{\prime}=3.29\times10^{15}$~Hz), which, from a model of a quasar placed within an exponential gas disk, is just
sufficient to ionise all of the neutral gas in a large spiral \citep{cw12}.  This explains the ubiquitous absence of
21-cm absorption in any source with $Q_\text{\HI}\gapp3\times10^{56}$~s$^{-1}$, at any redshift and independent of
source classification.
\citet{cww+08} suggested a selection effect, where the traditional pre-selection of targets of known optical redshifts biases
21-cm surveys towards objects that are the most UV luminous in the source rest-frame.
If the gas is completely ionised, even the Square Kilometre Array (SKA) will fail to detect 21-cm absorption
in the currently known high redshift radio sources,  with future searches being required to dispense  with the 
reliance upon an optical redshift to which to tune the receiver \citep{cwt+12,msc+15}. Nevertheless, in order to further test this and
improve the statistics, we have embarked upon a survey of a new sample of flat spectrum 
radio sources over a large redshift space. In \citet{chj+17} we reported our $z\gapp2.6$
survey with the Green Bank Telescope (GBT) and the Giant Metrewave Radio Telescope (GMRT) and
here we report our $z\lapp0.4$ GMRT observations.

\section{Source selection, observations and data reduction}
\label{sec:obs}

As in the previous stage of this survey \citep{chj+17}, the sources were selected from the {\em Second Realization of
  the International Celestial Reference Frame by Very Long Baseline Interferometry} (ICRF2, \citealt{mab+09}). This
comprises a sample of strong flat spectrum radio sources, of which 1682 now have known redshifts (\citealt{tm09,tsj+13}
and references therein), yielding a frequency to which to tune the receiver in the search for 21-cm absorption.
Furthermore, all are VLBI calibrators and so have significant compact flux, thus maximising the chance of a high
covering factor and thus optical depth (\citealt{cag+13}, see Sect, \ref{sec:or}). The goal of the survey was to search
and quantify the incidence of associated \HI\ 21-cm absorption over all redshifts and since our previous observations
\citep{chj+17} searched 19 sources at $z \gapp2.6$, we wished to complement this with lower redshift data. Band-5 of the
Upgraded Giant Metrewave Radio Telescope (uGMRT) spans a range of 1000--1450 MHz and, of the ICRF2 sources for which
1420 MHz is redshifted into this range (\HI\ 21-cm at $z\lapp0.42$), there are 24 for which the estimated flux density
exceeded $S_{\rm obs}\approx0.2$~Jy at the redshifted 21-cm frequency. Of these, there were 23 with sufficient
optical/UV photometry to determine the ionising photon rate, of which time was awarded to observe 19 (see Table
\ref{obs}).

In order to strike a balance between sensitivity and sample size, the $S_{\rm obs}\geq0.2$~Jy sources were observed for
a total of one hour each, including overheads (calibration and slewing).  The observations were taken with the full 30
antenna array on 23 March 2018, in two orthogonal circular polarisations (LL \& RR). For bandpass calibration 3C\,48,
3C\,147 and 3C\,298 were used, with the phases being self calibrated.  The Band-5 receiver was used with the GSB
back-end, over a bandwidth of 16 MHz and 512 channels, giving a spectral resolution of $\approx8$~\kms\ over $\pm2000$
\kms, in order to cover any uncertainties in the redshift. The data were calibrated and flagged using the {\sc miriad}
interferometry reduction package.  After averaging the two polarisations, a spectrum was extracted from the cube. At
these frequencies, radio frequency interference (RFI) was low and a satisfactory image was produced in all but five
cases\footnote{None of the sources was resolved by the synthesised beam, which ranged from $3.2\arcsec\times3.0\arcsec$
  to $5.6\arcsec\times4.5\arcsec$.}  (see Table \ref{obs}).

\section{Results}
\label{sec:res}
\subsection{Observational results} 

\label{sec:or}
\begin{figure*} 
\vspace{16.6cm}  
\includegraphics{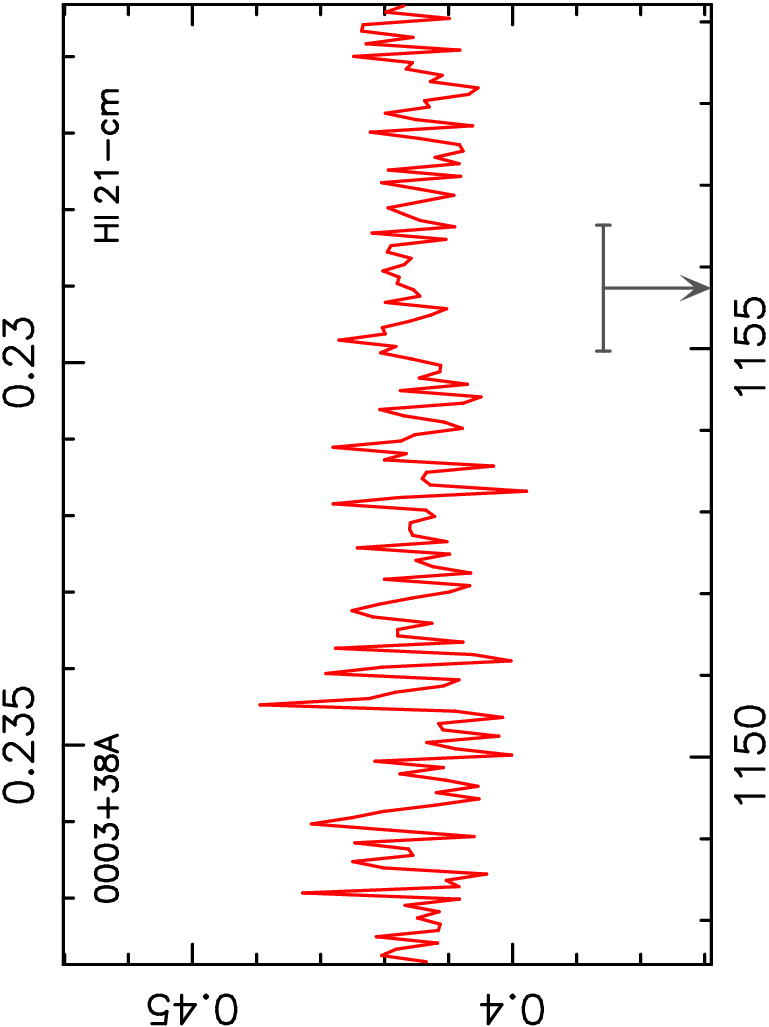}  
\includegraphics{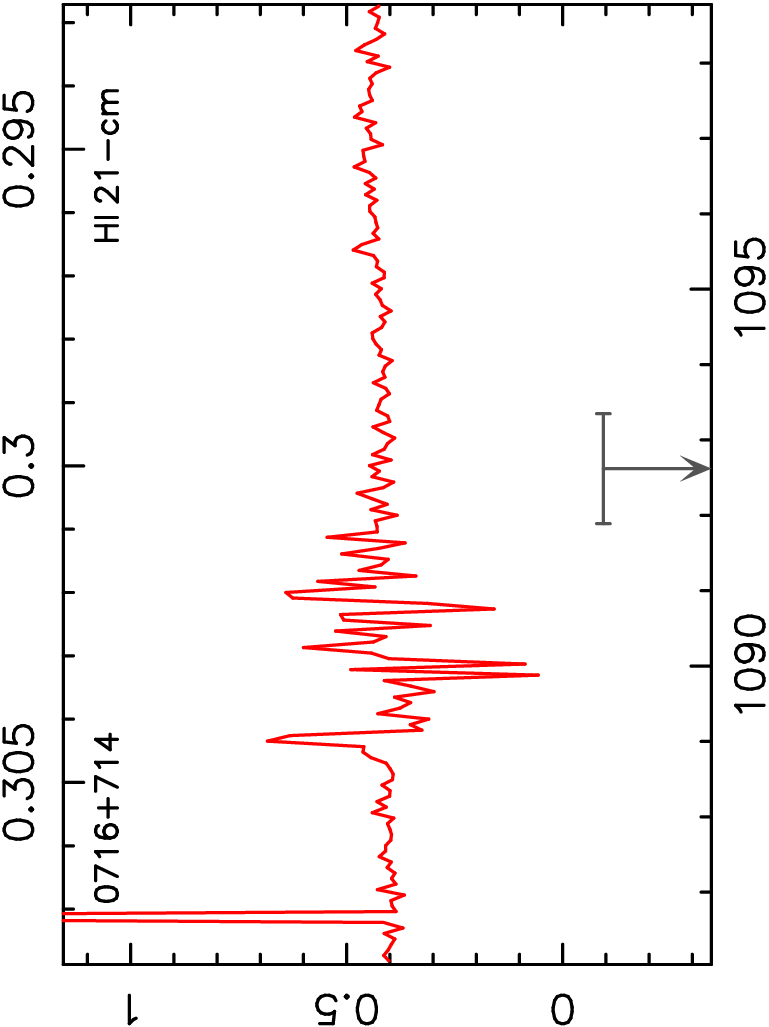} 
\includegraphics{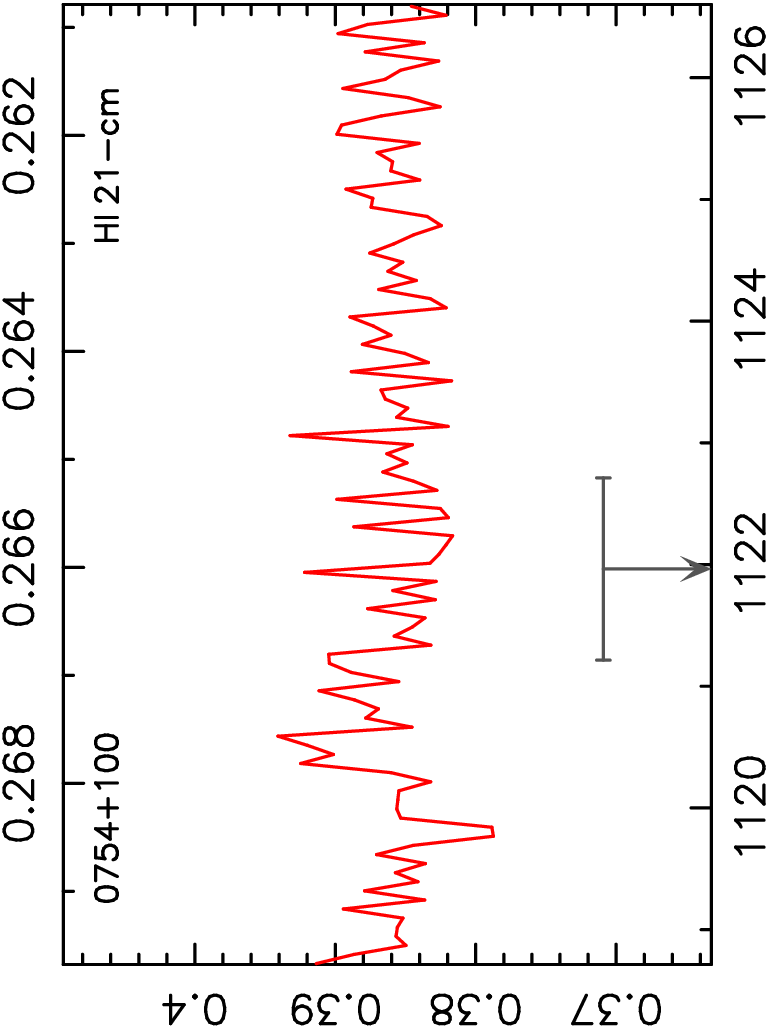} 
\includegraphics{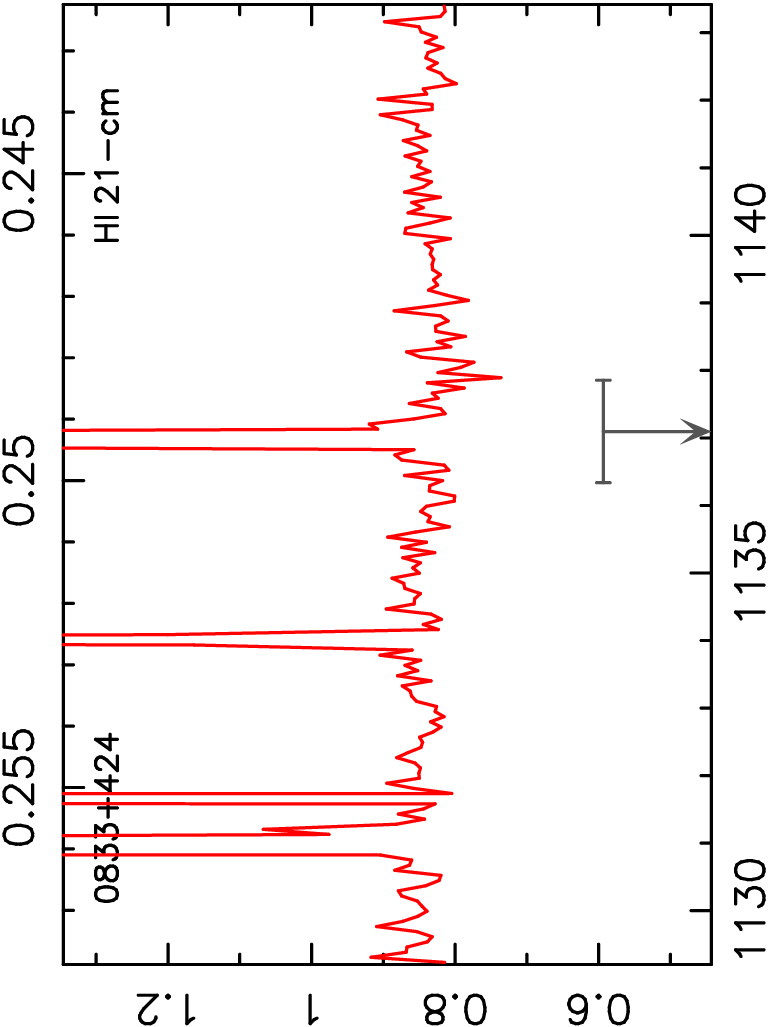}
\includegraphics{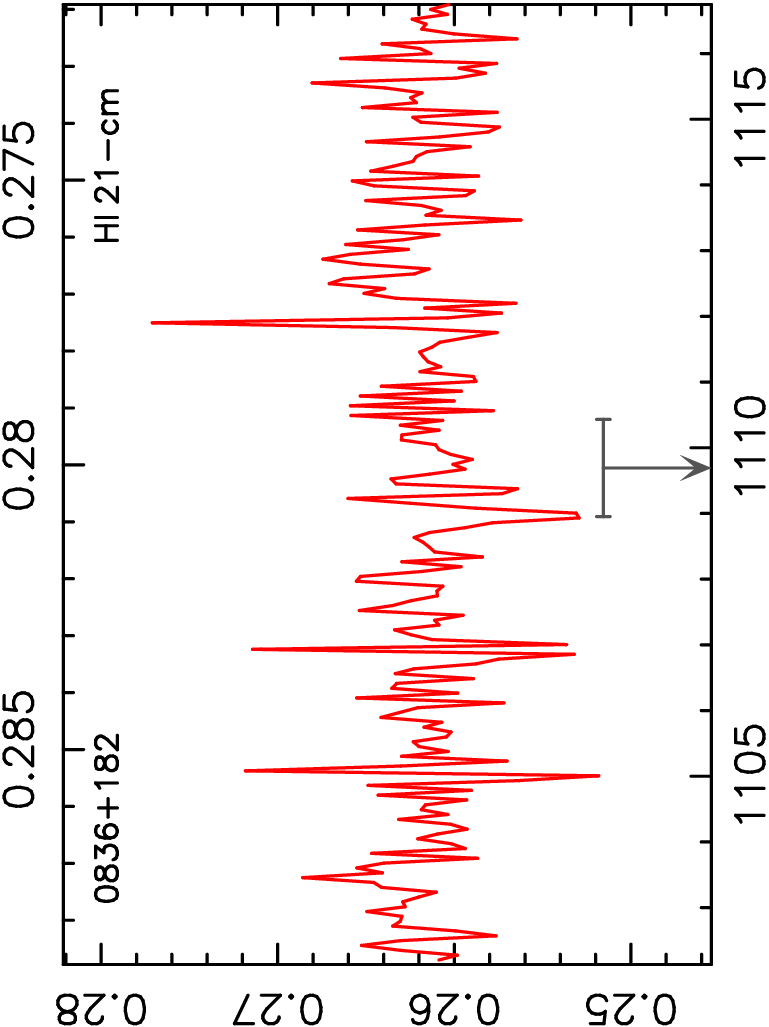}   
\includegraphics{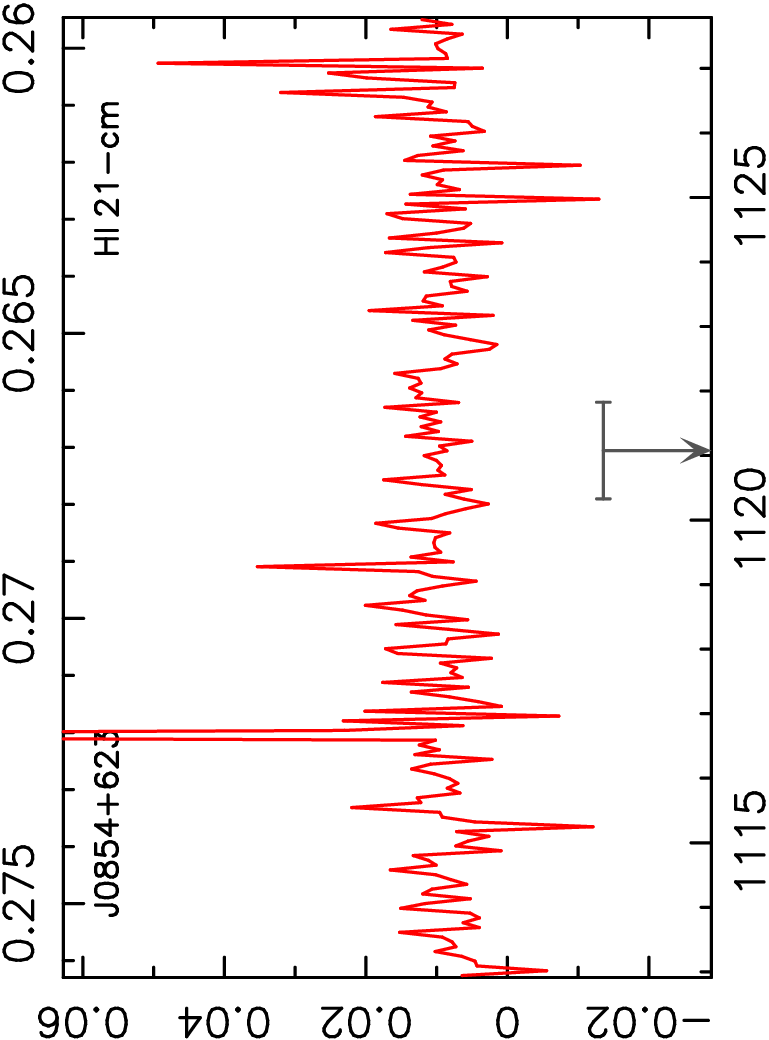} 
\includegraphics{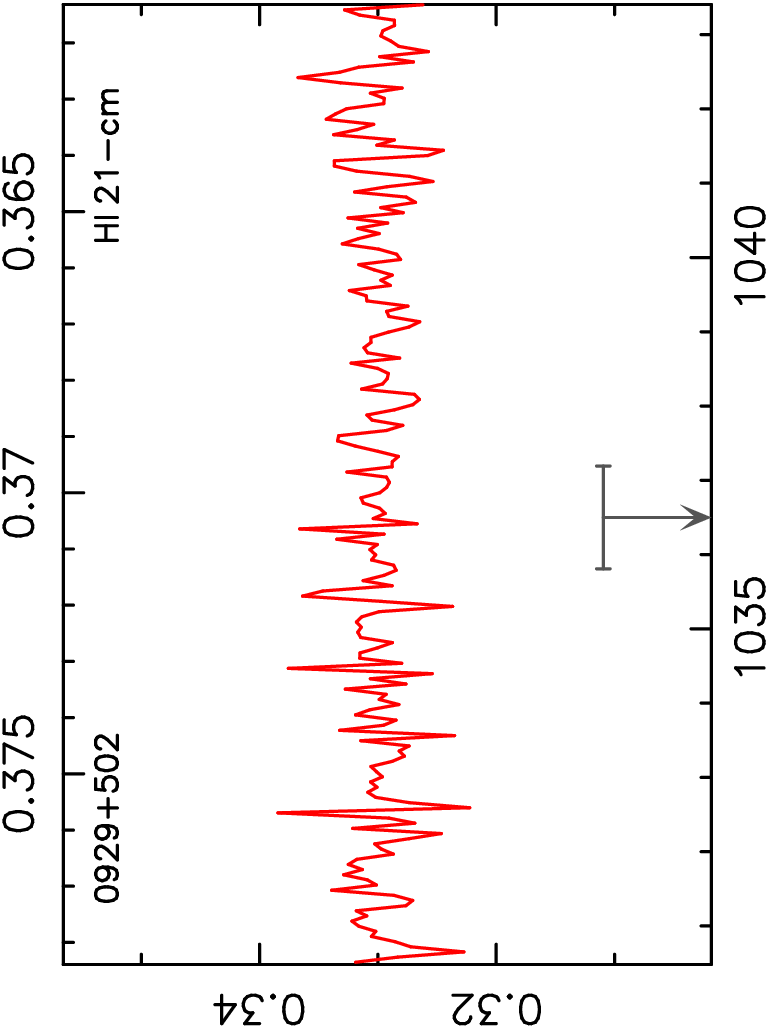} 
\includegraphics{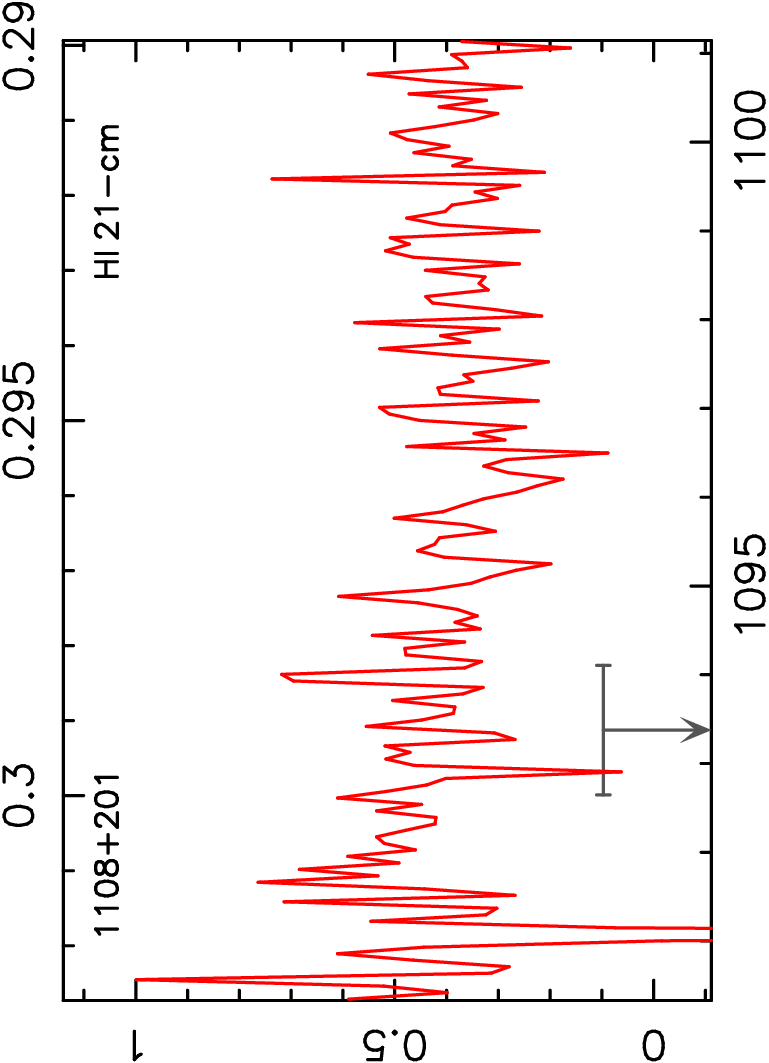}
\includegraphics{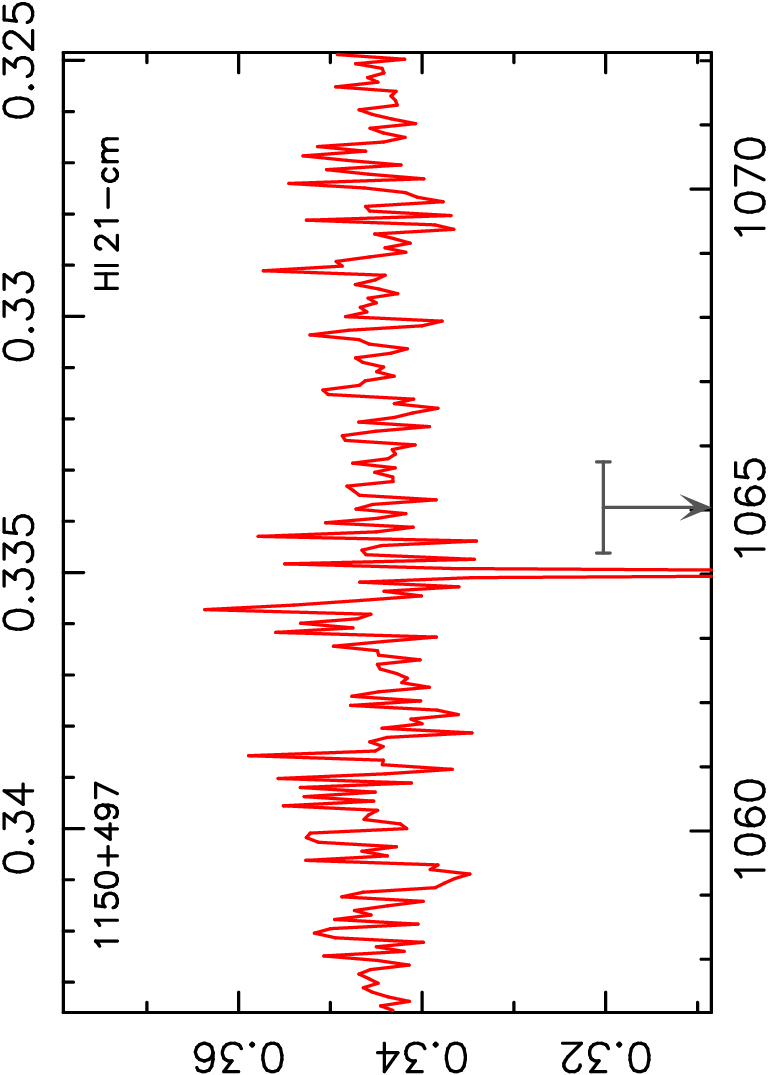}  
\includegraphics{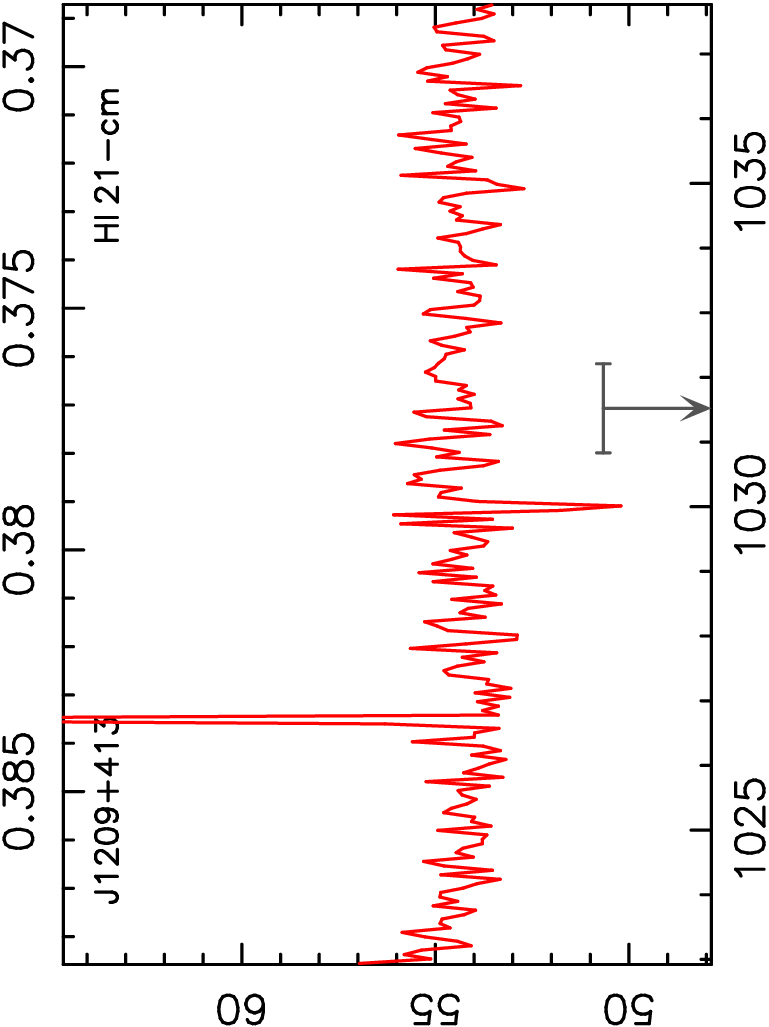}  
\includegraphics{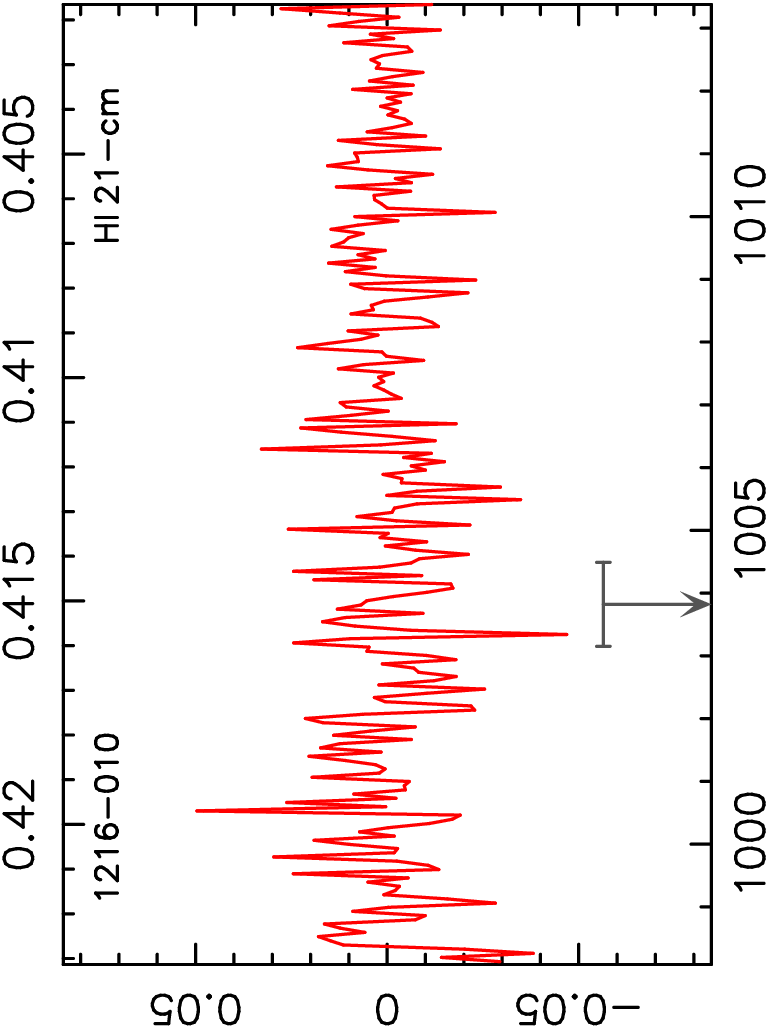} 
\includegraphics{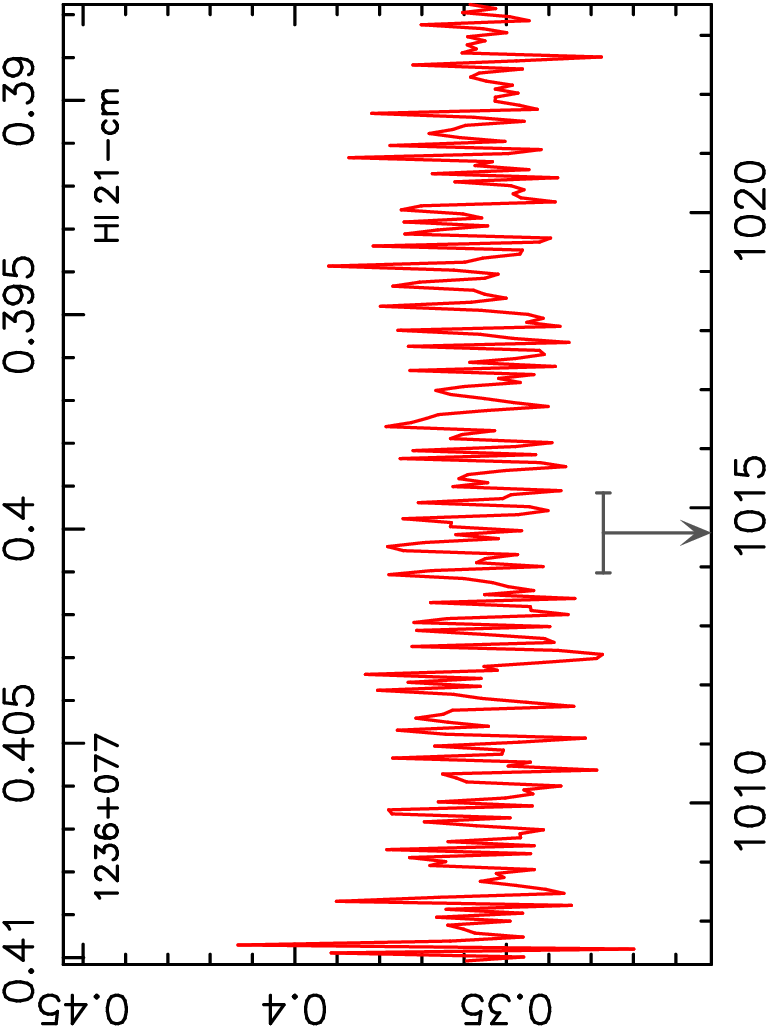}
\includegraphics{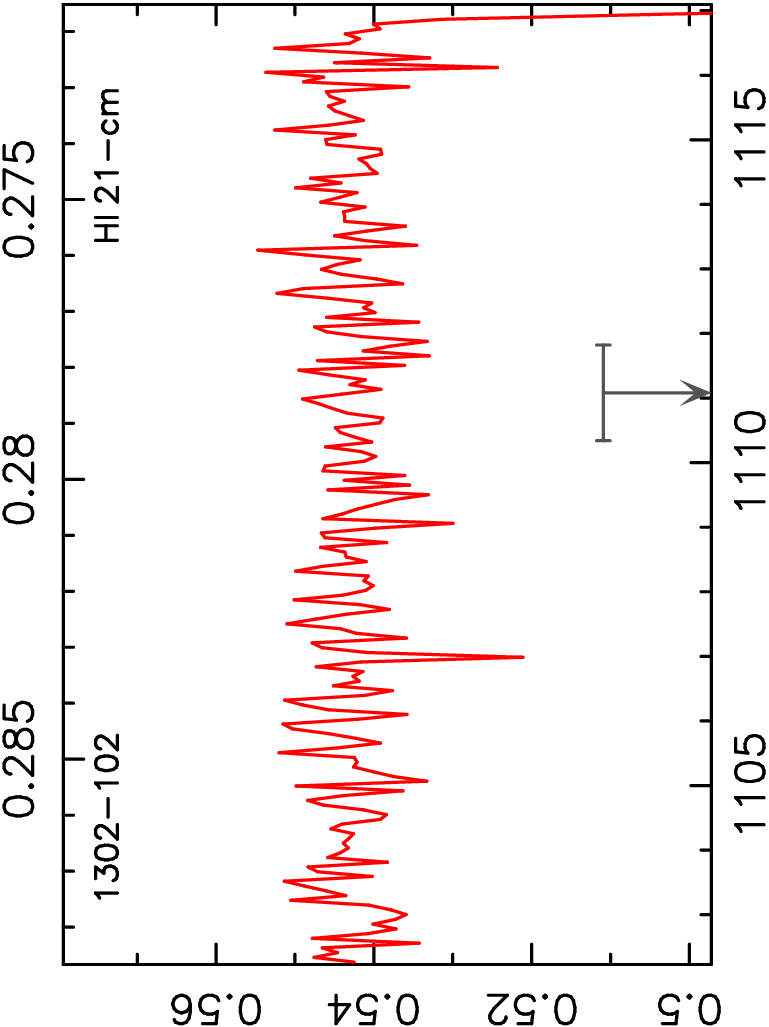} 
\includegraphics{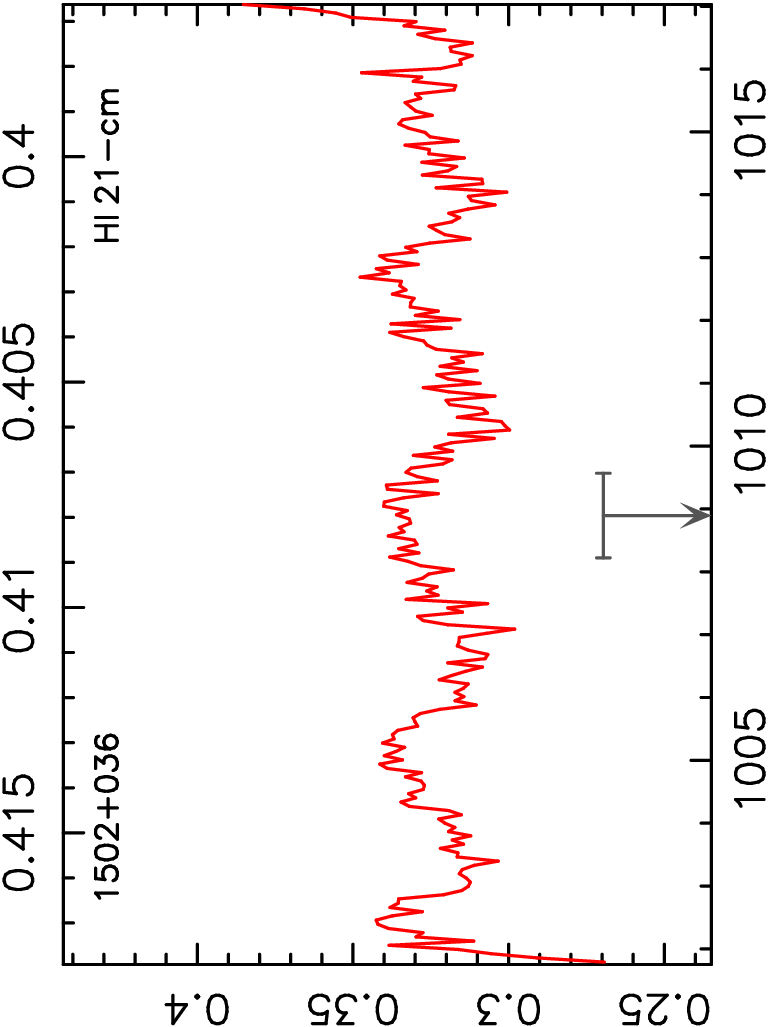}  
\includegraphics{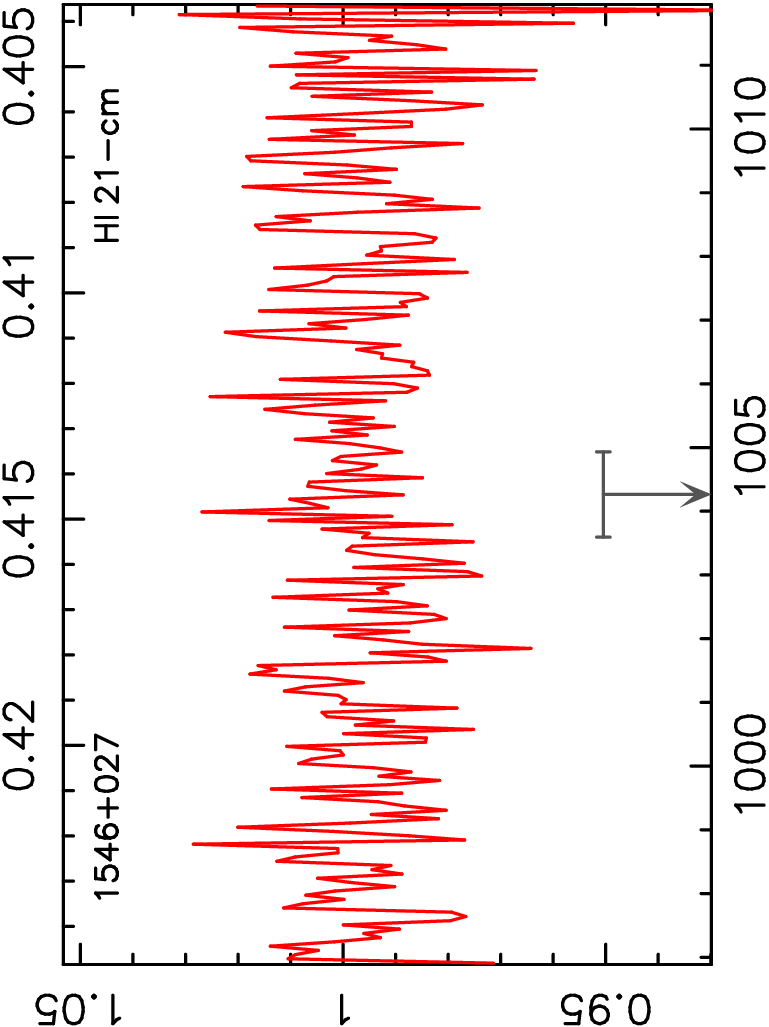} 
\includegraphics{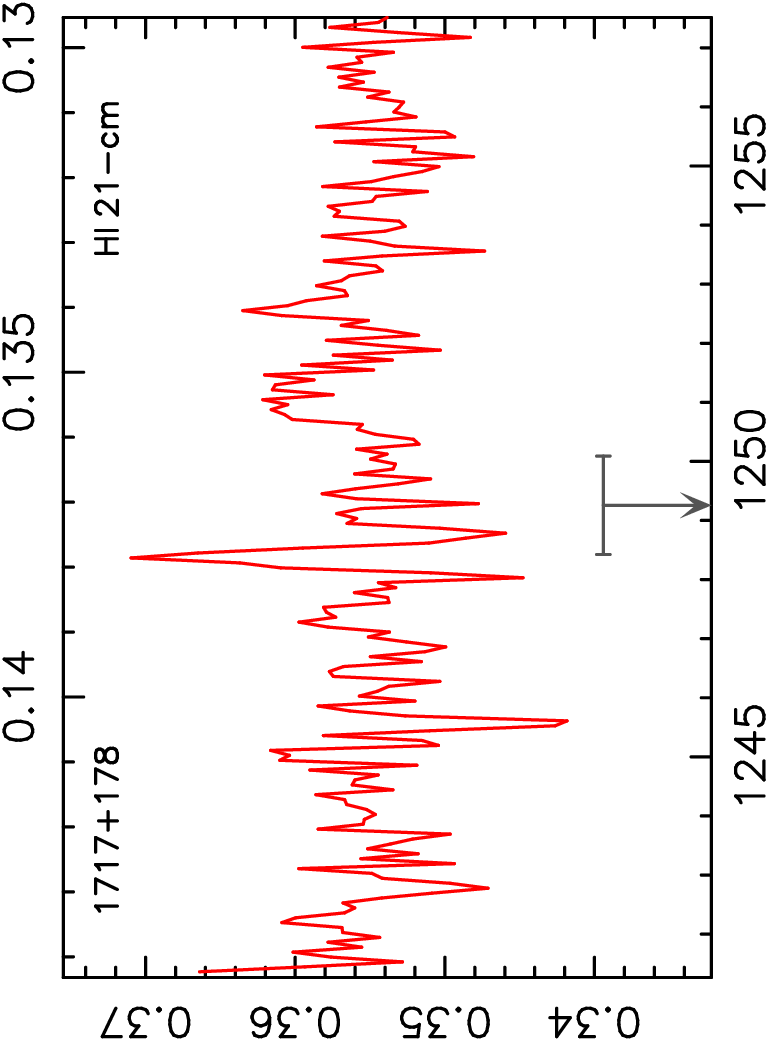} 
\includegraphics{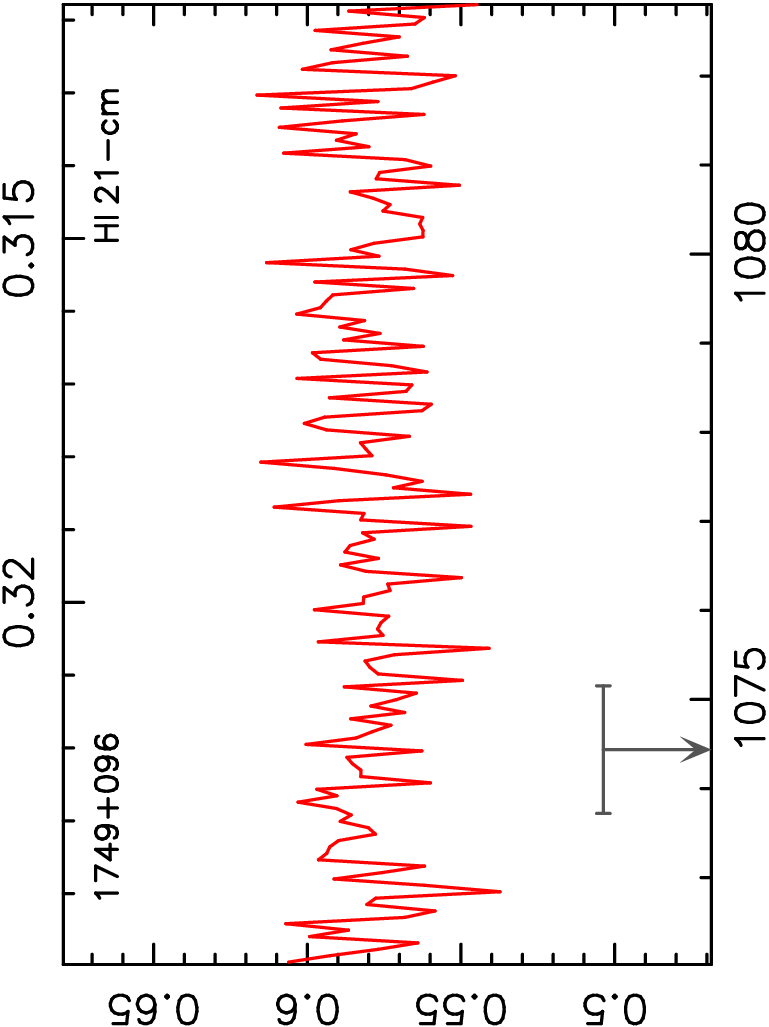}  
\includegraphics{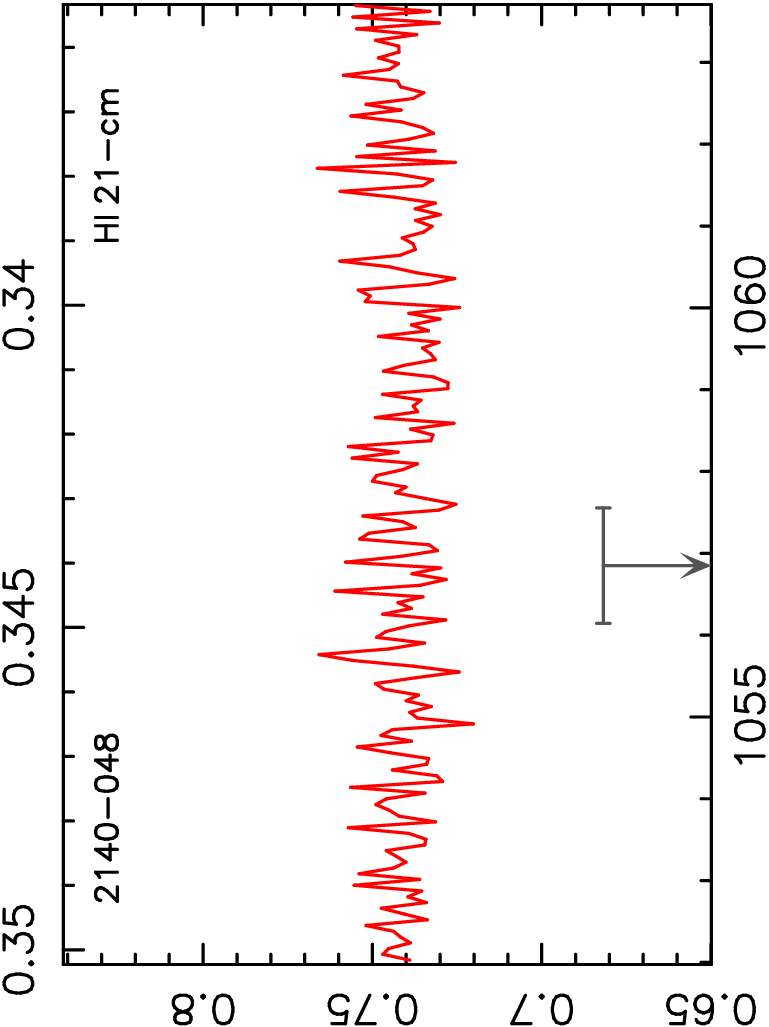} 
\includegraphics{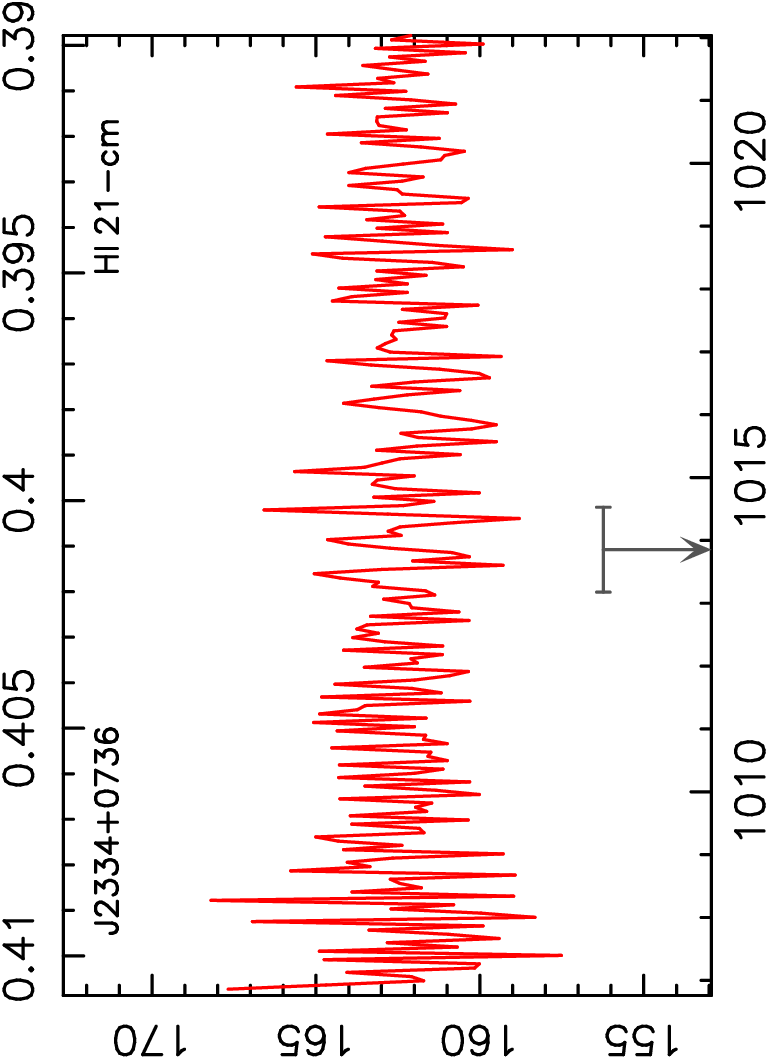} 
\caption{The reduced spectra shown at a spectral resolution of 20 \kms.  The ordinate gives the flux density [Jy], except
  for J1209+413 \& J2334+0738 (see Table \ref{obs}), and the abscissa the barycentric frequency [MHz]. The scale along the
  top shows the redshift of \HI\ 21-cm over the frequency range and the downwards arrow shows the frequency of
  the absorption  expected  from the optical redshift, with the horizontal bar showing a span of $\pm200$ \kms\ for guidance. Note that the features in 1150+497 and J1209+413 are also apparent off source in the image and so are not considered to be real.}
\label{spectra}
\end{figure*}

\begin{figure*}
\centering 
\includegraphics[angle=270,scale=0.62]{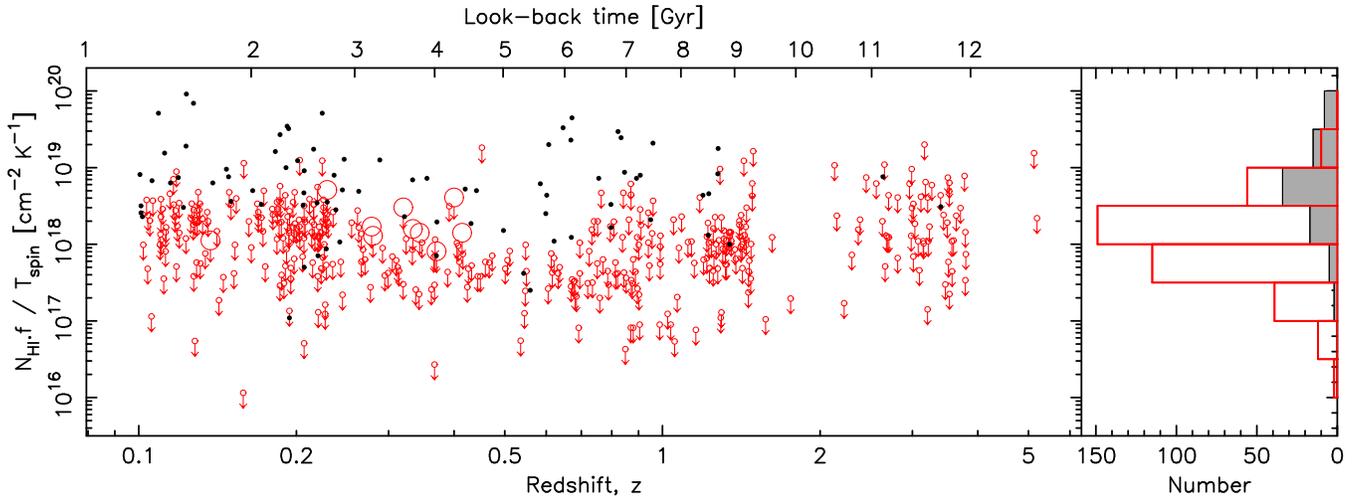}
\caption{The absorption strength [$1.823\times10^{18}\,(T_{\rm spin}/f) \int\!\tau\,dv$] versus redshift for the $z\geq0.1$
  associated \HI\ 21-cm absorption searches. The filled circles/histogram represent the detections and the unfilled
  circles/histogram the $3\sigma$ upper limits to the non-detections, with the large circles designating the new targets (Table~\ref{obs}).}
\label{N-z}
\end{figure*}

In Fig. \ref{spectra} we show the final spectra, smoothed to a spectral resolution of  $\Delta v=20$ \kms, and summarise  the results in Table~\ref{obs}.
\begin{table*}   
\centering
\begin{minipage}{174mm}
  \caption{The observational results. $z$ is the optical redshift of the source, $\Delta S$ the rms noise reached per 20
    \kms\ channel, $S_{\rm meas}$ is the measured flux density, $\tau_{3\sigma}$ the derived limit to the optical depth,
    where $\tau_{3\sigma}=-\ln(1-3\Delta S/S_{\rm meas})$ is quoted for these non-detections. These give the quoted
    column densities [\scm], where $T_{\rm spin}$ is the spin temperature and $f$ the covering factor, followed by the
    frequency [MHz] and redshift ranges over which the limit applies (between the RFI spikes to either side of the expected redshift).}
\begin{tabular}{@{}l c c  c l c r  c  @{}} 
\hline
\smallskip
Name     &  $z$ & $\Delta S$ [mJy] & $S_{\rm meas}$ [Jy] & $\tau_{3\sigma}$ & $N_{\text{\HI}}$  [\scm]  & $\nu$-range [MHz] & $z$-range \\
\hline
B2\,0003+38A   & 0.229 &  7.5  & 0.419 &$<0.054$   & $<2.0\times10^{18}\,T_{\rm spin}/f$ &  1147.51--1159.10 &  0.22543 --0.23781\\  
\protect[HB89]\,0716+714  & 0.300 & 20.7 &  0.434 &\multicolumn{2}{c}{\sc rfi dominant}   & 1086.83--1098.54 & 0.29299--0.30693\\
\protect[HB89]\,0754+100 & 0.266&  2.94 & 0.386   &  $<0.023$ & $<8.3\times10^{17}\,T_{\rm spin}/f$& 1118.81--1126.47 & 0.26094--0.26957 \\
FBQS\,J083353.8+422401 & 0.249153 &  --- & --- & \multicolumn{2}{c}{\sc rfi dominant}  & 1137.53--1143.13 & 0.24256--0.24868\\
\protect[HB89]\,0836+182 & 0.280 & 3.13 & 0.262 & $<0.036$ &  $<1.3\times10^{18}\,T_{\rm spin}/f$& 1102.33--1116.50 & 0.27219--0.28855 \\
WISE\,J085450.56+621850.1 & 0.267 & --- & --- & \multicolumn{2}{c}{\sc rfi dominant}  & 1113.21--1127.59  & 0.25968--0.27596\\
WISE J092915.43+501336.0 & 0.370387 & 2.47 & 0.330 & $<0.022$ & $<8.2\times10^{17}\,T_{\rm spin}/f$& 1030.78--1043.15 & 0.36166--0.37799 \\ 
PKS\,1108+201 & 0.2991 &  --- & --- & \multicolumn{2}{c}{\sc rfi dominant}  &  1090.47--1100.93 & 0.29019--0.30256\\
SBS\,1150+497  & 0.33366 & 4.91 & 0.345 & $<0.043$& $<1.6\times10^{18}\,T_{\rm spin}/f$& 1064.50--1071.88 & 0.32516--0.33434\\
FBQS\,J120922.7+411941  & 0.377 &--- & ---  & \multicolumn{2}{c}{\sc no flux calibration} & 1023.36--1037.52 & 0.36904--0.38798 \\
\protect[HB89]\,1216--010  &  0.415 & 13.4 & --- &   \multicolumn{2}{c}{\sc insufficient flux}& 998.13--1013.15  & 0.40197-- 0.42307\\
\protect[HB89]\,1236+077 & 0.400 &  13.3 & 0.356 &  $<0.11$ & $<4.1\times10^{18}\,T_{\rm spin}/f$ &  1007.62--1023.33 &0.38802--0.40967 \\
PG\,1302--102 & 0.2784 &  8.28 & 0.543&  $<0.046$&  $<1.7\times10^{18}\,T_{\rm spin}/f$& 1102.38--1116.52 & 0.27217--0.28849\\
\protect[HB89]\,1502+036 & 0.40788 &  -- & 0.324& \multicolumn{2}{c}{\sc bandpass ripple dominant}& 1002.11--1016.80  &  0.39694--0.41742 \\
\protect[HB89]\,1546+027  & 0.41438 & 13.2 & 0.997 & $<0.040$ & $<1.4\times10^{18}\,T_{\rm spin}/f$& 997.11--1011.65 & 0.40405--0.42452 \\
\protect[HB89]\,1717+178 & 0.137 & 3.63 & 0.355 & $<0.031$ & $<1.1\times10^{18}\,T_{\rm spin}/f$& 1246.80--1258.66 & 0.12851--0.13924 \\
\protect[HB89]\,1749+096 & 0.322 & 15.7 & 0.580& $<0.081$& $<3.0\times10^{18}\,T_{\rm spin}/f$& 1241.37--1257.21 & 0.12981--0.14422\\
\protect[HB89]\,2140--048  & 0.344 & 9.12 & 0.741 &  $<0.037$& $<1.4\times10^{18}\,T_{\rm spin}/f$& 1052.13--1063.65 & 0.33541--0.35003\\
WISE\,J233412.82+073627.6 & 0.401 & --- &  --- & \multicolumn{2}{c}{\sc no flux calibration}& 1006.88--1021.98&  0.38985--0.41070\\
\hline
\end{tabular}
{Notes: 0003+38A detected at $N_{\text{\HI}} = 3.5\times10^{18}(T_{\rm spin}/f)$  \scm\ by \citet{ak18}, 0754+100 observed to a sensitivity of  $< 7.3\times10^{17}(T_{\rm spin}/f)$  \scm\ per 30 \kms\ channel  by \citet{gdb+15}, 1108+201 to $<8.1\times10^{18}(T_{\rm spin}/f)$ \scm\ per 17.8 \kms\ channel by \citet{ak17}. Flux calibration was not possible for 1206+416 nor J2334+0736,
with the task {\sf gpboot} over-scaling the fluxes.}
\label{obs}  
\end{minipage}
\end{table*} 
The velocity integrated optical depth of the 21-cm absorption, $\int\!\tau\,dv$, is related
to the total neutral hydrogen column density  via  
\begin{equation}
N_{\rm HI}  =1.823\times10^{18}\,T_{\rm  spin}\int\!\tau\,dv,
\label{enew_full}
\end{equation}
where $T_{\rm spin}$ is the spin temperature of the gas, which is a measure of the excitation from the lower hyperfine
level \citep{pf56,fie59,be69}. 
The observed optical depth, $\tau_{\rm obs}$, is the ratio of the line depth, $\Delta S$, to the observed background flux, $S_{\rm obs}$, and is
related to the intrinsic optical depth via
\begin{equation}
\tau \equiv-\ln\left(1-\frac{\tau_{\rm obs}}{f}\right) \approx  \frac{\tau_{\rm obs}}{f}, {\rm ~for~}  \tau_{\rm obs}\equiv\frac{\Delta S}{S_{\rm obs}}\lapp0.3,
\label{tau_obs}
\end{equation}
where the covering factor, $f$, is the fraction of $S_{\rm obs}$ intercepted by the absorber.
Therefore, in the optically thin regime (where $\tau_{\rm obs}\lapp0.3$), Equ. \ref{enew_full} can be approximated as
\begin{equation}
N_{\rm HI}  \approx 1.823\times10^{18}\,\frac{T_{\rm  spin}}{f}\int\!\tau_{\rm obs}\,dv, 
\label{enew}
\end{equation}
which we use to derive the absorption strength sensitivities.  Of the 19 targets, the absorption strength could be obtained for 11
which we show in comparison to the previously published values (Fig.~\ref{N-z}).\footnote{These are compiled from
  \citet{dsm85,mir89,vke+89,ubc91,cps92,cmr+98,cwh+07,mcm98,ptc99,ptf+00,rdpm99,mot+01,ida03,vpt+03,cwm+06,cww+08,cwm+10,cwwa11,cwsb12,cwt+12,caw+16,cwa+17,gss+06,omd06,kpec09,ems+10,ssm+10,css11,cgs13,ace+12,asm+15,ysdh12,ysd+16,gmmo14,sgmv15,akk16,akp+17,ak17,ak18,chj+17,gdb+15,jgs18,adi19}. Note
  that, in order to compare our limits with those in the literature, each has been re-sampled to the same spectral
  resolution (20 \kms, as in Fig. \ref{spectra}). This is used as the FWHM to obtain the integrated optical depth
  limit, thus giving the $N_{\text \HI} f/T_{\rm spin}$ limit per channel (see
  \citealt{cur12}).}
Our sensitivities span from $N_{\text{\HI}} = 10^{17.9} - 10^{18.7} T_{\rm spin}/f$~\scm\ per 20~\kms\ channel, which is 
within the range of the previous detections at $z\geq0.1$.\footnote{We restrict our analysis to these redshifts, due to possible weakening of the  absorption by coincident 21-cm emission \citep{cd18}.}  

Note that, although we reach an rms noise level of 18~mJy per 8.41 \kms\ (giving a 
$3\sigma$ a sensitivity to $N_{\text{\HI}}> 2.0\times10^{18}\,T_{\rm spin}/f$~\scm) 
in 0003+38A, we do not detect the $N_{\text{\HI}}=3.5\times10^{18}\,T_{\rm spin}/f$~\scm\ absorption of \citet{ak18} in the
cube extracted spectrum. In the {\em uv} data, however, a feature is apparent in the RR polarisation only, with a
similar full-width at zero intensity of FWZI\,$\approx50$~\kms, although the feature is shallower (15, cf. 40 mJy) and
offset at $+1750$ \kms\ (cf. $-50$ \kms) from $z=0.2290$ (i.e. at $z=0.2362$).

\subsection{Ionising photon rates}
\label{sec:ipr}

Following the  procedure  of \citet{cwsb12}, 
for each source we obtained the photometry from {\em NASA/IPAC Extragalactic
  Database} (NED), the {\em Wide-Field Infrared Survey Explorer} (WISE, \citealt{wem+10}), {\em Two Micron All Sky
  Survey} (2MASS, \citealt{scs+06}) and the {\em Galaxy Evolution Explorer} (GALEX, data release
GR6/7)\footnote{http://galex.stsci.edu/GR6/\#mission} databases. Each flux datum, $S_{\nu}$,  was converted to a luminosity, via
$L_{\nu}=4\pi \, D_{\rm L}^2 \,S_{\nu}/(z+1)$, where $D_{\rm L}$ is the luminosity
 distance to the source, and corrected for Galactic extinction
\citep{sfd98}. A power-law was then fit to the UV rest-frame data, allowing the ionising photon rate, $Q_\text{\HI}\equiv
\int^{\infty}_{\nu}({L_{\nu}}/{h\nu})\,d{\nu}$, to be derived. 

The derived ionsing photon rates shown in comparison to the previously published searches in Fig. \ref{Q-z}.\footnote{The spectral energy distributions (SEDs) of our targets are shown in Fig.~3 of the published version.} 
\addtocounter{figure}{+1}    
\begin{figure*}
\centering 
\includegraphics[angle=270,scale=0.62]{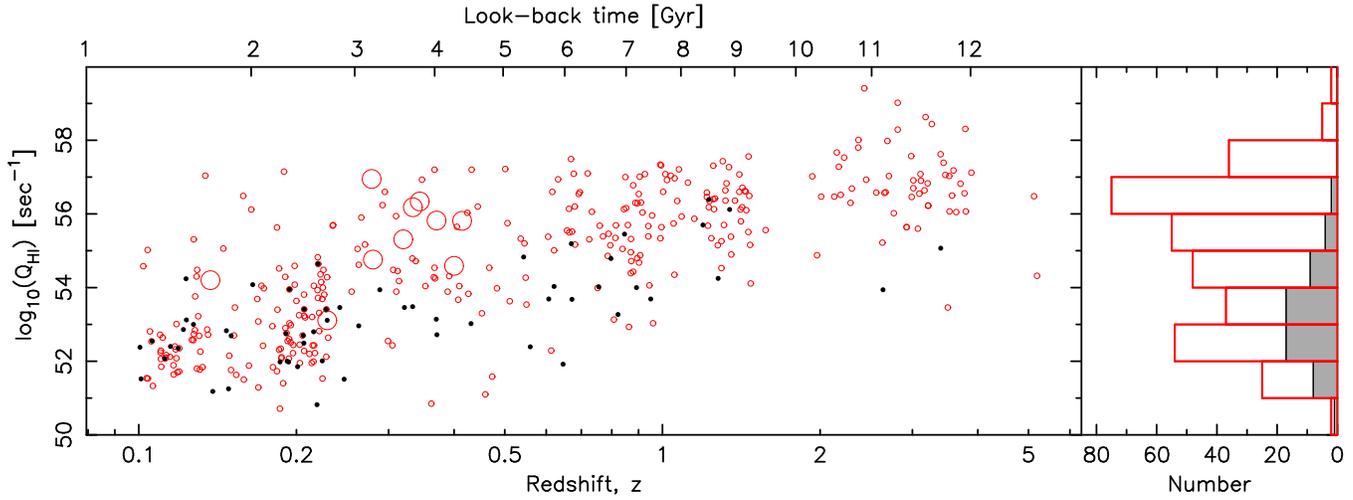}
\caption{The ionising ($\lambda \leq912$ \AA) photon rate versus redshift for the \HI\ 21-cm absorption searches.  The
  symbols and histogram are as per Fig. \ref{N-z}.}
\label{Q-z}
\end{figure*} 
Of the 397 $z\geq0.1$ sources with sufficient UV photometry, there are 58 detections and 252 non-detections with
ionising photon rates up to the highest value where 21-cm absorption has been detected
($Q_\text{\HI}=2.45\times10^{56}$~s$^{-1}$ in PKS\,1200+045, \citealt{ak17}).\footnote{Note that \citet{akp+17} detect
  21-cm absorption at $z=1.223$ in TXS\,1954+513 for which they claim $L_{\rm UV}\approx4\times10^{23}$ \WpHz. This,
  however, is extrapolated from just two optical ($R$-band and $B$-band) photometry points, which we deem insufficient (see
  \citealt{cwsb12}).}  Applying this 18.7\% detection rate to the $Q_\text{\HI}> 2.45\times10^{56}$~s$^{-1}$ sources,
gives a binomial probability of $1.49\times10^{-8}$ of obtaining 0 detections out of 87 searches. This is significant at
$5.66\sigma$, assuming Gaussian statistics, and supports the finding that the lack of detections at high redshift is due
to the optical pre-selection of targets biasing towards sources which are sufficiently UV luminous to ionise all the gas
in the host \citep{cw12}.

The ionising photon rates for the targets,  not dominated by RFI, are listed in Table~\ref{Q-table}.
\begin{table}  
 \caption{The ionising photon rates for the targets for which we could obtain limits to the 21-cm absorption strength (Table \ref{obs}).}
\begin{tabular}{@{}l c r  c   @{}} 
\hline
\smallskip
Source     &  $z$ & $N_{\text{\HI}}$  [\scm]   & $Q_\text{\HI}$ [s$^{-1}]$ \\
\hline
0003+38A   & 0.229 & $3.5\times10^{18}\,T_{\rm spin}/f$ & $1.3\times10^{53}$\\
0754+100 & 0.266 & $<5.9\times10^{17}\,T_{\rm spin}/f$& $7.9\times10^{55}$\\  
0836+182 & 0.280 & $<1.3\times10^{18}\,T_{\rm spin}/f$&  $5.8\times10^{54}$\\ 
J0929+5013& 0.370387 & $<8.2\times10^{17}\,T_{\rm spin}/f$&  $6.6\times10^{55}$\\
1108+201 & 0.2991 &   $<4.2\times10^{17}\,T_{\rm spin}/f$    & $3.6\times10^{52}$\\
1150+497  & 0.33366 & $<1.6\times10^{18}\,T_{\rm spin}/f$& $1.5\times10^{56}$\\
1236+077 & 0.400 & $<4.1\times10^{18}\,T_{\rm spin}/f$ & $3.9\times10^{54}$\\
1302--102 & 0.2784 & $<1.7\times10^{18}\,T_{\rm spin}/f$& $8.9\times10^{56}$\\
1546+027  & 0.41438 & $<1.4\times10^{18}\,T_{\rm spin}/f$ & $6.5\times10^{55}$\\
1717+178 & 0.137 & $<1.1\times10^{18}\,T_{\rm spin}/f$& $1.6\times10^{54}$\\
1749+096 & 0.322 & $<3.0\times10^{18}\,T_{\rm spin}/f$ & $2.0\times10^{55}$\\
2140--048  & 0.344 &$<1.4\times10^{18}\,T_{\rm spin}/f$& $2.2\times10^{56}$\\
\hline
\end{tabular}
{Notes: $N_{\text{\HI}} = 3.5\times10^{18}(T_{\rm spin}/f)$ \scm\ in 0003+38A from \citet{ak18}, $<5.9\times10^{17}(T_{\rm spin}/f)$ \scm\ in 0754+100 from \citet{gdb+15} and  $<4.2\times10^{17}(T_{\rm spin}/f)$ \scm\  in 1108+201 from \citet{ak17}, upon re-sampling to a spectra resolution of 20 \kms.}
\label{Q-table}  
\end{table} 
The obtained sensitivities span $N_{\text{\HI}} = 0.6 - 3.0\times10^{18}\,T_{\rm spin}/f$ \scm. At $N_{\text{\HI}} >3.0\times10^{18}\,T_{\rm spin}/f$, there are 65 detections and 122 non-detections  below the critical ionising rate, giving a detection rate of 35\%.
For the 11 targets with $Q_\text{\HI}\lapp3\times10^{56}$~s$^{-1}$, we therefore expect $3.9\pm2.0$ detections where
one is obtained (by \citealt{ak18}), and so the detection rate is not especially anomalous. This is confirmed by applying the
global 18.7\% probability of detection below the critical luminosity, which gives a binomial probability of 0.3621 of obtaining
one or fewer detections out of 11 searches (significant at just $0.91\sigma$).

Given that the sample is larger with more comprehensive photometry\footnote{Giving 397 compared to 211 sources
  previously \citep{chj+17}.}, we now we examine other selection factors, over and above the ionising photon rate,
which could affect the 21-cm absorption detection rate (e.g. the degree of dust reddening and the coverage of the background flux, \citealt{cwa+17}).

\section{Discussion}

\subsection{Photo-ionisation versus excitation by 21-cm photons}
\label{sec:photo}

\begin{figure}
\centering 
\includegraphics[angle=270,scale=0.57]{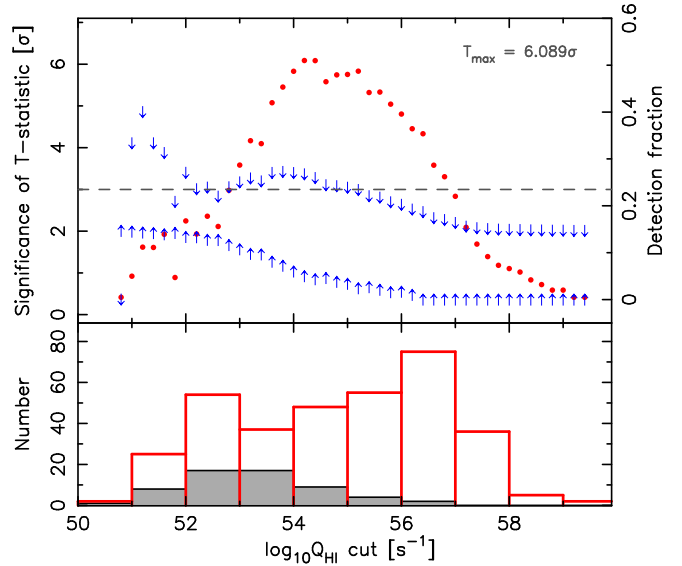}
\caption{The $T$-statistic of the difference in the proportions, $\hat{p}_1$ and $\hat{p}_2$, for the detection of \HI\
  21-cm absorption at various cuts of the ionising photon rate. The downward/upwards arrows show the
  detection fraction (right hand scale) below/above the cut (proportions $\hat{p}_1$ and $\hat{p}_2$, respectively).  
The significance is seen to drop to $<3\sigma$ at $Q_\text{\HI}\gapp10^{57}$ ~s$^{-1}$, due to the 
lack of detections ---  shown in the bottom panel, where the filled histogram shows the number of 
detections and the unfilled the number of non-detections. }  
\label{T-Q}
\end{figure}

Firstly, we examine the suggestion of \citet{akk16,ak17,ak18} that gas excitation by the incident radio
continuum may be responsible for the paucity of 21-cm detections at high redshift.  
As mentioned in Sect.~\ref{sec:or}, the spin temperature of the gas can be raised by excitation to the upper hyperfine level by 21-cm photons
and the Malmquist bias would cause the high redshift sources to be the most luminous. Due to the detection of 21-cm absorption over all of the same
radio luminosities as the non-detections, this was deemed an unlikely cause by \citet{cww+08}, with \citet{cw10} 
using the $T$-statistic to show that the UV was dominant over the radio luminosity. 
Since the sample has increased significantly,  however,  we revisit this. The $T$-statistic is given by 
\[ 
T = \frac{\hat{p}_1 - \hat{p}_2}{\sqrt{\hat{p}(1-\hat{p})(N_1^{-1}+N_2^{-1})}},
\] 
where $\hat{p}_1=X_1/N_1$ and $\hat{p} _2=X_2/N_2$ are the two
measured proportions  (i.e. the detection rate below and above the cut) and $\hat{p}=(X_1+X_2)/(N_1+N_2)$ is the total
proportion. This has the standard normal distribution under the null hypothesis that the proportions
($\hat{p}_1$ and $\hat{p}_2$) are the same, which we reject for $Q_\text{\HI}\gapp10^{53}$ ~s$^{-1}$, where the
difference between the two proportions is significant at $>3\sigma$ (Fig. \ref{T-Q}).

\begin{figure}
\centering 
\includegraphics[angle=270,scale=0.57]{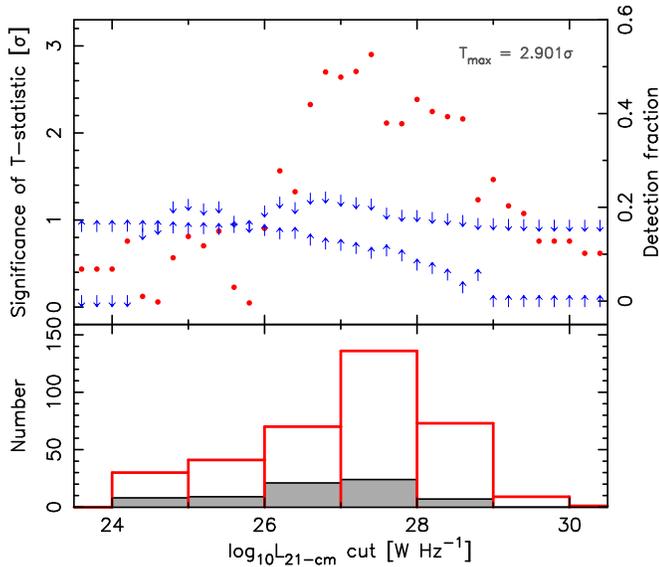}
\caption{As Fig. \ref{T-Q}, but for the 21-cm continuum luminosity.}  
\label{T-21}
\end{figure}

In Fig. \ref{T-21} we show the $T$-statistic for the rest-frame 1420 MHz continuum luminosity, obtained
from the second order polynomial fit to the radio SEDs. 
This peaks at $<3\sigma$, indicating that excitation by the continuum radio luminosity has less of an effect on the
detection of 21-cm absorption. The increase at $L_{\rm 21-cm} \gapp 10^{26}$ \WpHz\  is most likely due
to the correlation between the radio and UV luminosities, which are both degenerate with redshift \citep{cw10}.

\begin{figure}
\centering 
\includegraphics[angle=270,scale=0.52]{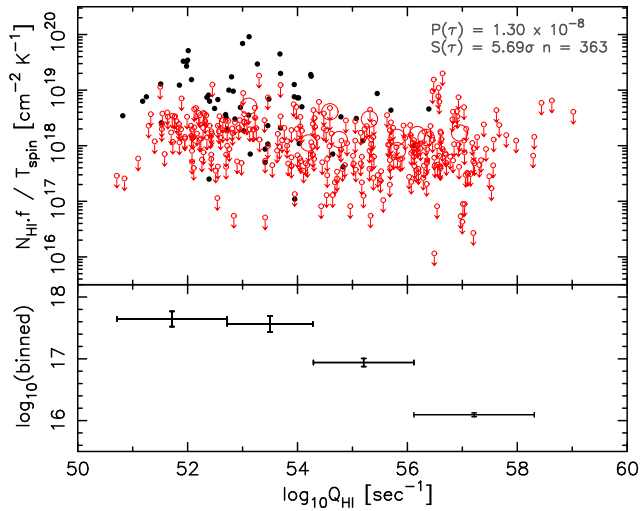}
\caption{The absorption strength versus the ionising photon rate.  In the bottom panel, the data are shown in equally sized
  bins, where the horizontal bars show the range of points in the bin and the vertical error bars the $1\sigma$
  uncertainty in the mean value.  The limits are incorporated, via the Kaplan--Meier estimator, which gives a
  maximum-likelihood estimate based upon the parent population \citep{fn85}.}
\label{N-Q}
\end{figure}

\begin{figure}
\centering 
\includegraphics[angle=270,scale=0.52]{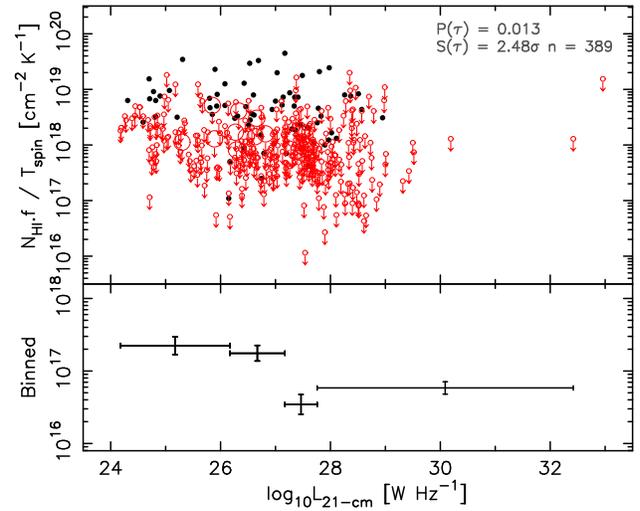}
\caption{The absorption strength versus the 21-cm continuum luminosity.}
\label{N-21}
\end{figure}

Examining this further, by including the limits to the absorption strength via the {\em Astronomy SURVival Analysis}
({\sc asurv}) package \citep{ifn86}, a generalised non-parametric Kendall-tau test gives a probability of $P(\tau) =
1.30\times10^{-8}$ of the $N_{\text{\HI}}f/T_{\rm spin} -Q_\text{\HI}$ anti-correlation occuring by chance
(Fig. \ref{N-Q}),
which  is significant at $S(\tau) = 5.69\sigma$. 
For the continuum radio luminosity, the correlation remains weak (Fig. \ref{N-21}), thus 
suggesting that photo-ionisation of the neutral gas is the primary reason for the non-detection of 21-cm absorption
in the high redshift sources.

\subsection{Other factors}

\subsubsection{Source reddening}

\begin{figure}
\centering 
\includegraphics[angle=270,scale=0.57]{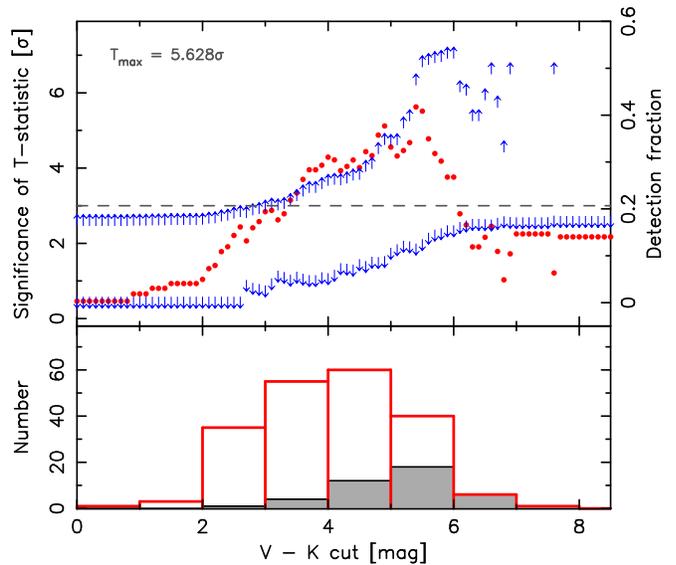}
\caption{As Fig. \ref{T-Q}, but for the visible--near-infrared colour.}
\label{T-VminusK}
\end{figure}

Another factor that is shown to have an effect on the \HI\ 21-cm detection rate is the redness of the source
\citep{wfp+95,cmr+98}, with \citet{cwm+06,cwa+17,cw10} finding a correlation between visible (or blue)--near-infrared
colour and the absorption strength.  Although the visible magnitudes will be degenerate with the ionising photon rate,
this may provide evidence of dust shielding the neutral gas from ambient UV radiation.  This effect has also been
observed for the molecular absorption strength in intervening absorption systems \citep{cwc+11}, where the absorbing gas
is remote from the ionsing UV continuum, suggesting that the reddening is indeed at least partially due to dust.  This,
however, has been recently disputed by \citet{ak18}, who found no correlation between the absorption strength and degree
of reddening (through the red--near-infrared colour).  Obtaining the visible and near-infrared magnitudes from the
photometry fits of the entire sample (Sect.~\ref{sec:ipr}), in Fig.~\ref{T-VminusK} we show the $T$-statistic for the
visible--near-infrared colour,
from which we see that the degree of reddening has a significant effect on the detection rate  at $V-K\gapp3$. Furthermore, 
\begin{figure}
\centering 
\includegraphics[angle=270,scale=0.52]{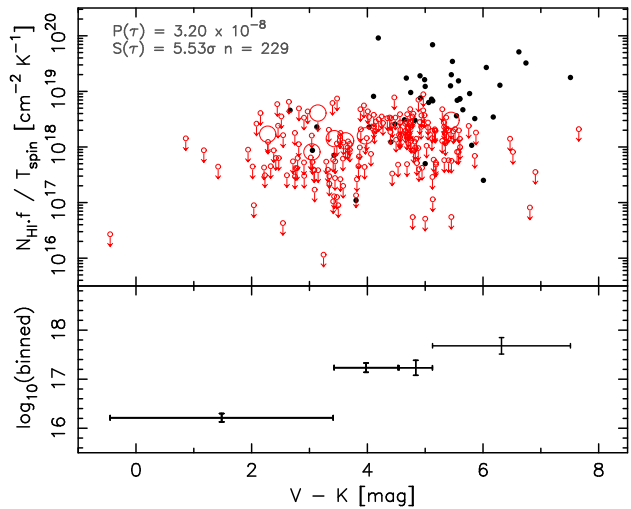}
\caption{The absorption strength versus the visible--near-infrared colour.}
\label{V-K}
\end{figure} 
Fig. \ref{V-K}  demonstrates a strong correlation between the absorption strength and the degree of reddening (and
also suggests that our targets may not be sufficiently reddened). This is inconsistent with the finding of 
 \citet{ak18}, who use $R-K$ -- the $R$-magnitude will not  be as susceptible to dust attenuation as the $V$-magnitude.
Additionally,  they use a smaller (58 cf. 230 sources, for which we could
obtain the magnitudes) and more biased (containing $\alpha_{\rm 21-cm}> -0.5$ sources only) sample.

\subsubsection{Spectral index}

In addition to excitation of the gas raising the spin temperature, the coverage of the continuum flux can affect the
detection of 21-cm absorption: Low coverage ($f <1$) will reduce the observed optical depth (Equ. \ref{tau_obs}), as has
been observed in the case of intervening absorption \citep{cmp+03}. For associated absorption, the observed optical
depth is known to be anti-correlated with the extent of the radio emission, consistent with $f\propto \tau_{\rm obs}$
\citep{cag+13}. 

Generally, the radio emission extents are unknown and the extent of the absorbing medium would also be required to
determine the covering factor. One proxy for the covering factor, via the radio source size, is the spectral
index. Here, extended radio sources beaming along our sight-line are expected to have flatter radio SEDs than those in
which the lobes are projected in the plane of the sky \citep{ffs+90}. Indeed, \citet{cwsb12} attributed their
non-detections in eight $L_{\rm UV} \lapp 10^{23}$ \WpHz\ sources to the selection of ultra-steep spectrum ($\alpha_{\rm
  21-cm}\lapp-1$) sources at $z\gapp3$ \citep{dvs+02}.  Therefore, any correlation between the spectral index and
absorption strength would suggest a strong relationship between the spectral slope and the extent of the radio emission.
\begin{figure}
\centering 
\includegraphics[angle=270,scale=0.57]{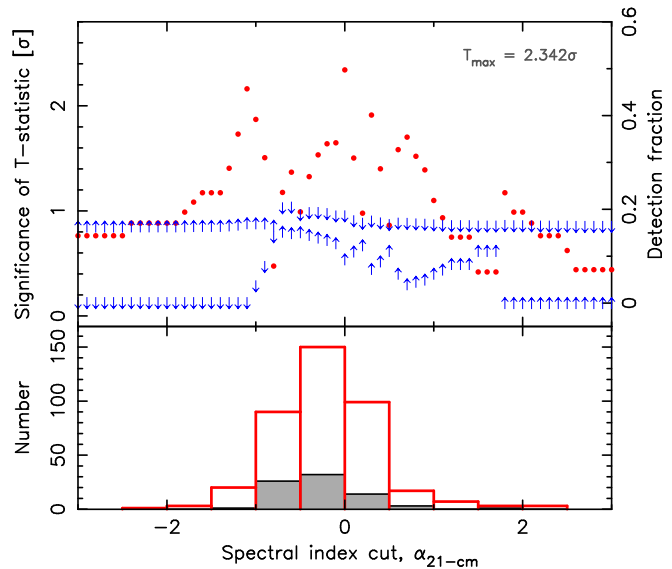}
\caption{As Fig. \ref{T-Q}, but for the rest-frame 1420 MHz spectral index (see Sect \ref{sec:photo}).}  
\label{T-SI}
\end{figure} 
From the $T$-statistic (Fig. \ref{T-SI}), we do see that the detections may favour flat spectrum ($\alpha_{\rm 21-cm}\approx0$)  sources,
although this is not significant and so could be the result of small numbers at $|\alpha_{\rm 21-cm}|\gapp1$.

\subsubsection{Turnover frequency}

Another tracer of radio source size is the presence of a turnover in the SED, due to free--free absorption/synchrotron self-absorption 
at low frequencies causing a turnover at $\nu_{_{\rm TO}}\sim$~GHz. 
In such gigahertz peaked sources (GPS), the turnover frequency is anti-correlated with the source size 
(e.g. \citealt{ode98,fan00,omd06}) and a possible correlation between the detection of 21-cm absorption and the turnover frequency  was suggested in near-by
galaxies, where $\left<\nu_{_{\rm TO}}\right> = 10^{8.66\pm0.32}$ Hz for the detections, compared to $\left <\nu_{_{\rm TO}}\right> = 10^{8.00\pm0.21}$~Hz for the non-detections \citep{cras16}

\begin{figure}
\centering 
\includegraphics[angle=270,scale=0.57]{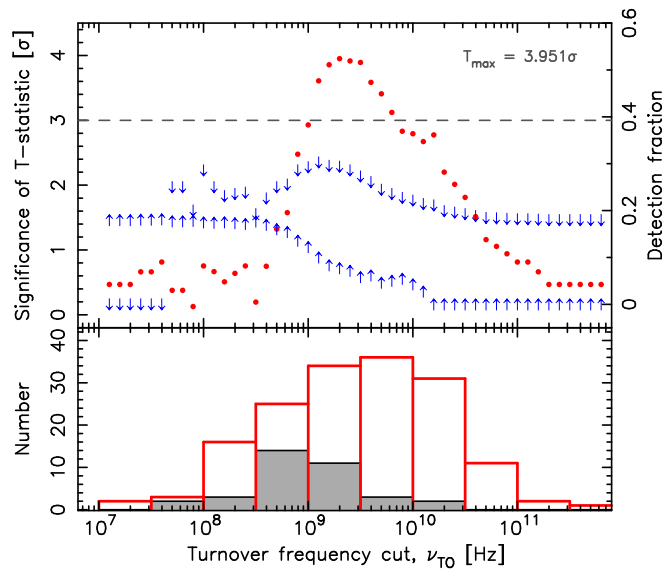}
\caption{As Fig. \ref{T-Q}, but for the radio-band turnover frequency.}
\label{T-TO}
\end{figure}

\begin{figure}
\centering 
\includegraphics[angle=270,scale=0.52]{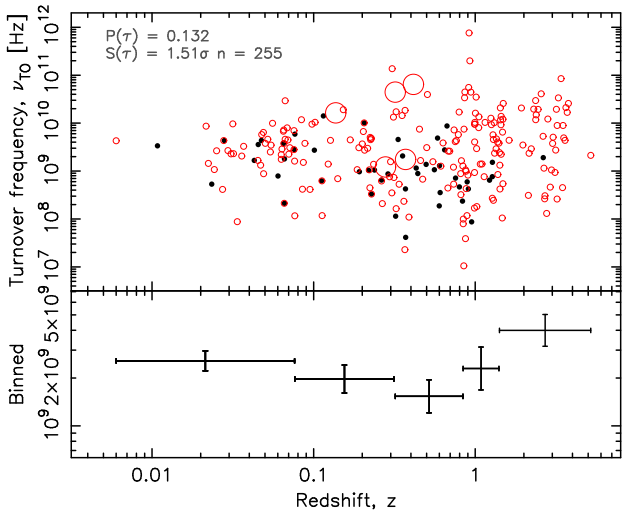}
\caption{The rest-frame turnover frequency versus the redshift for all of the published \HI\ 21-cm absorption searches which
exhibit a turnover.}
\label{TO-z}
\end{figure}

\begin{figure}
\centering 
\includegraphics[angle=270,scale=0.52]{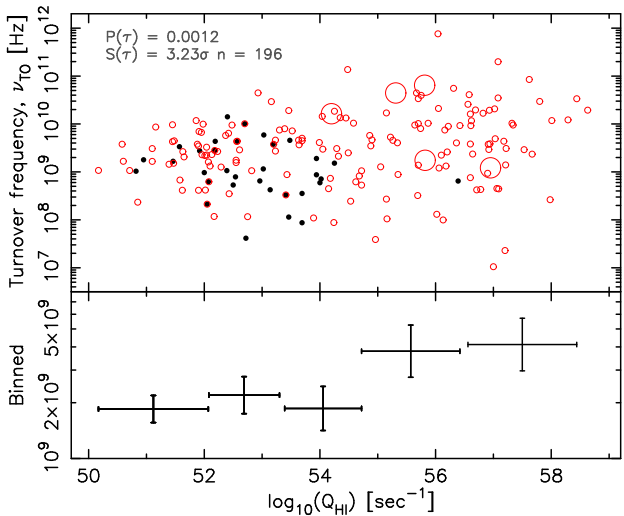}
\caption{As Fig. \ref{TO-z}, but for the ionising photon rate.}
\label{TO-Q}
\end{figure}

\begin{figure}
\centering 
\includegraphics[angle=270,scale=0.52]{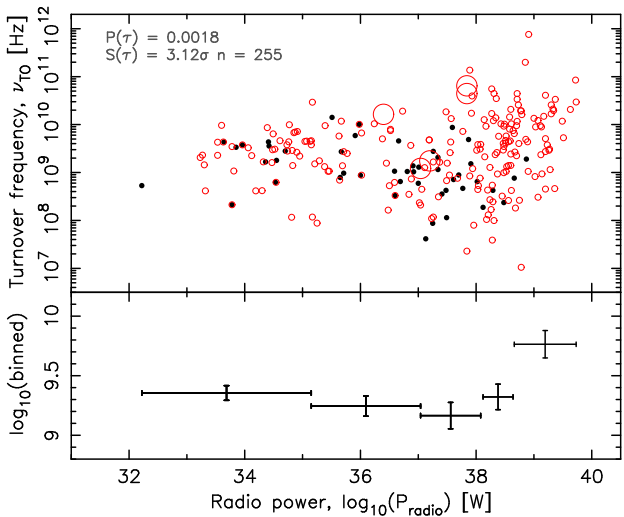}
\caption{As Fig. \ref{TO-z}, but for the radio power.}
\label{TO-P}
\end{figure}

For the whole sample, we obtain the rest-frame (intrinsic) turnover frequency from the second order polynomial fit to the 
radio SEDs (see \citealt{cwsb12}),
which shows a significant ($>3\sigma$) effect for  $\nu_{_{\rm TO}}\gapp1$~GHz (Fig.~\ref{T-TO}). 
However, this is due to a {\em lower} 21-cm detection rate at higher values of $\nu_{_{\rm TO}}$, which would suggest that the detection rate decreases
with the background continuum size, contrary to what we expect from $f\propto \tau_{\rm obs}$.
\citet{dbo97} showed that the turnover frequency increases with redshift, which they attributed to the higher radio luminosities
increasing the electron density.
Investigating this, we find only a weak $\nu_{_{\rm TO}}-z$ correlation (Fig.~\ref{TO-z}), 
which may  be driven by the ionsing photon rate (Fig. \ref{TO-Q}), rather
than the radio power (Fig. \ref{TO-P}).
This is consistent with photo-ionisation of the gas increasing the electron density  to such
an extent that it negates any detection benefit of the smaller source sizes.

\section{Conclusions}

As part of our ongoing survey to determine the incidence of associated \HI\ 21-cm absorption with redshift, we report on
the now complete low redshift phase of this. We have observed 19 compact radio sources at $z\lapp0.4$, for which we
obtain useful spectra for 11. Three of the sources have been previously searched, with one being detected
\citep{ak17,ak18,gdb+15}.  Combining these with our useful spectra, gives 11 sources which have ionising photon rates
below the critical ionsing photon rate, from which we expect four detections, based upon the general detection rate over
the sensitivities reached. Given that the statistics have significantly increased, due to a larger sample and more
complete photometry, we revisit the main observational factors believed to affect the detection of \HI\ 21-cm
absorption:
\begin{enumerate}

\item We confirm that the photo-ionisation of the neutral gas by high ionising  ($\lambda \leq912$ \AA) photon rates 
is most likely responsible for the paucity of 21-cm absorption at high redshift. The binomial probability of the exclusive
non detections found at ionising photon rates of $Q_\text{\HI}\gapp3\times10^{56}$ ~s$^{-1}$ is significant
at $5.66\sigma$, consistent with hypothesis that all of the neutral gas in the host is ionised above these rates \citep{cw12}.

\item The difference between the detection and non-detection fractions reaches a significance of $6.09\sigma$ for
  $Q_\text{\HI}$, whereas this peaks at $2.90\sigma$ for the radio luminosity. This confirms the finding of
  \citet{cww+08,cw12} that the non-detection of neutral atomic gas at high redshift is dominated by photo-ionisation of
  the gas, rather than excitation to the upper hyper-fine level by 1420 MHz photons \citep{akk16,ak17,ak18}.

\item We confirm that the reddening of the source is correlated with the strength of the 21-cm absorption
  \citep{cwm+06,cwa+17,cw10}, with a $5.53\sigma$ correlation between visible--near-infrared colour and absorption
  strength. A low degree of reddening will, of course, be degenerate with the faint observed $V$ magnitudes in the UV
  luminous sources, although the fact that this is also seen for intervening absorbers \citep{cwc+11} suggests that
  at least some of the reddening is due to the presence of dust. 

\item That the detection rate shows no strong dependence on the radio-band spectral index, which is believed
to trace the projected extent of the background source size and thus the coverage of the flux
by the absorber. However, the majority of the sources are flat spectrum with $|\alpha_{\rm 21-cm}|\lapp0.5$. 

\item Since the source size is anti-correlated with radio turnover frequency for gigahertz peaked sources,
we also expect a correlation with the detection rate. We do find a strong relationship, although this is 
in the opposite sense to that expected with the detection rate decreasing with $\nu_{_{\rm TO}}$, where
the source sizes are believed to be smaller (\citealt{ode98,fan00,omd06}). This is, however,
consistent with the turnover frequency increasing with redshift, although it is possible that the main
driver is the higher ionising photon rate rather than higher radio power \citep{dbo97}. 
It does, nevertheless, support their argument that higher turnover frequencies are a tracer of higher
electron densities in gigahertz peaked sources.
\end{enumerate}
Thus, we conclude that photo-ionisation of the gas is the main factor in rendering detections of \HI\ 21-cm absorption
rare at redshifts of $z\gapp1$: A high redshift source sufficiently bright to provide a spectrum in the optical band is
extremely UV luminous in the source rest-frame, thus introducing a selection effect where the optical pre-selection of
targets biases against the sources most likely to contain large quantities of cool neutral gas. Again, this appears to
be independent of the radio properties, implying an ubiquitous effect, and reinforces our conclusion that the
pre-selection of targets based upon the optical spectra must be foregone in favour of wide-band radio observations.

\section*{Acknowledgements}

We wish to thank the anonymous referee for their prompt and helpful comments. We also thank the staff of the GMRT who
have made these observations possible. GMRT is run by the National Centre for Radio Astrophysics of the Tata Institute
of Fundamental Research.  The National Radio Astronomy Observatory is a facility of the National Science Foundation
operated under cooperative agreement by Associated Universities, Inc This research has made use of the NASA/IPAC
Extragalactic Database (NED) which is operated by the Jet Propulsion Laboratory, California Institute of Technology,
under contract with the National Aeronautics and Space Administration. This research has also made use of NASA's
Astrophysics Data System Bibliographic Services and {\sc asurv} Rev 1.2 \citep{lif92a}, which implements the methods
presented in \citet{ifn86}.


\label{lastpage}

\end{document}